\documentclass{amsart}

\usepackage{amssymb}
\usepackage{amsmath}
\usepackage{amsfonts}
\usepackage[foot]{amsaddr}
\usepackage{natbib}
\usepackage{graphicx}
\usepackage{placeins}
\usepackage{array}
\usepackage{multirow}
\usepackage{multicol}
\usepackage{dcolumn}
\usepackage{hyperref}
\usepackage{setspace}
\usepackage{tabularx}
\usepackage{booktabs}
\usepackage{float}
\usepackage[table,xcdraw]{xcolor}
\usepackage{grfext}
\usepackage{geometry}
\theoremstyle{definition}
\newtheorem{definition}{Definition}[section]

\newtheorem{proposition}{Proposition}[section]
\newtheorem{observation}{Observation}[section]
\newtheorem{model}{Model}[section]
\theoremstyle{remark}
\newtheorem{remark}{Remark}[section]
\theoremstyle{plain}

\newtheorem{notation}{Notation}[section]
\newtheorem{problem}{Problem}[section]

\numberwithin{equation}{section}

\DeclareMathOperator{\Unif}{Unif}
\DeclareMathOperator{\Dir}{Dir}
\DeclareMathOperator{\sgn}{sgn}

\usepackage{todonotes}
\newcommand{\dsnote}[1]{\todo[color=green!30, inline]{DS: #1}}

\let\rasterize=N

\ifx\rasterize Y
  \usepackage{epstopdf-base}
\else
\fi

\newcolumntype{R}{>{\raggedleft\arraybackslash}X}

\title[Nonparametric Detection of Gerrymandering...]{Nonparametric Detection of Gerrymandering in Multiparty Elections}
\author[W. Słomczyński]{Wojciech Słomczyński$^{1, 2}$}
\author[D. Stolicki]{Dariusz Stolicki$^{1, 3}$}
\author[S. Szufa]{Stanisław Szufa$^{1}$}
\date{March 2023}

\email{dariusz.stolicki@uj.edu.pl}
\email{wojciech.slomczynski@im.uj.edu.pl}

\address{$^{1}$ Jagiellonian Center for Quantitative Political Science, Jagiellonian University, ul. Reymonta 4, 31-007 Kraków, Poland.}
\address{$^{2}$ Institute of Mathematics, Jagiellonian University.}
\address{$^{3}$ Institute of Political Science, Jagiellonian University.}

\thanks{This research has been funded under the Polish National Center for Science grant no. 2019/35/B/HS5/03949 and supported by the QuantPol Center flagship project under the Jagiellonian University Excellence Initative, DigiWorld Research Area.}

\begin{document}

\maketitle

\onehalfspacing

\section{Preliminaries}

Most of the traditional methods developed for detecting gerrymandering in first-past-the-post electoral systems assume that there are only political parties really contesting the election, or, at least, that the party system is regular in the sense that all parties field candidates in every district. This is certainly a very reasonable assumption in many cases: under a well-known empirical regularity known as \emph{Durverger's law} FPTP tends to be correlated with the emergence of two-party systems. Moreover, many of the authors working on gerrymandering detection are motivated by U.S. legislative elections (state and federal), where the regular two-party pattern of competition prevails. However, in many other systems using FPTP we discover significant deviations from such patterns in the form of regional parties, strong independent candidates, minor parties that forgo campaigning in districts where there is no local party organization, etc. In the face of such deviations, many of the traditional methods fail completely, as we shall point out below. Our objective, therefore, is to develop a method of detecting gerrymandering that can be applied to such partially-contested multiparty election.

\subsection{Contribution}

Our main contribution consists of the development of a \emph{nonparametric methods for detecting gerrymandering} in \emph{partially-contested multiparty elections}. By \emph{nonparametric} we mean that, unlike most of the traditional statistical methods, the proposed method is free of assumptions about the probability distribution from which observed data points are drawn or the latent mechanism through which such data is generated. Instead we use statistical learning to identify regularities on the basis of the available empirical data.

By a \emph{partially-contested multiparty elections} we mean any FPTP election where at least some candidates are affiliated into one or more \emph{political parties} (after all, if every district-level election is completely independent and candidates cannot be affiliated into blocks, the very concept of gerrymandering as traditionally defined is meaningless), but \textit{for every party there is at most one affiliated candidate in every electoral district} (so there is no electoral intra-party competition). For the sake of simplicity, we treat independent (i.e., non-party-affiliated) candidates as singleton parties.

In particular, we permit the following deviations from the two-party pattern of competition:
\begin{itemize}
    \item the number of parties can differ from two,
    \item the number of candidates within each district can differ from two,
    \item a party can run candidates in any number of electoral districts,
    \item the set of parties contesting the election varies from one district to another.
\end{itemize}

Another area in which our approach differs from traditional methods for detecting gerrymandering is that they have been tailored towards \emph{testing a large ensemble of elections} (not necessarily from the same jurisdiction) rather than a single election. For instance, our original scenario was to test for evidence of gerrymandering in close to 2,500 Polish municipal elections. In particular, the proposed methods, like all statistical learning methods, require the researcher who wants to use them to have a large \emph{training set of elections} that they believe to be sufficiently similar insofar as the translation of votes into seats is concerned. If there is a large ensemble of elections being tested, they might form such a training set itself. There is no requirement that the training set and the tested set be disjoint as long as we can assume that gerrymandering is not ubiquitous in the testing set.

\subsection{Prior Work}


Among the methods of detecting gerrymandering that focus on the political characteristics of the districting plan (e.g., its impact on seats-votes translation or district-level vote distribution) the earliest focused on measuring how actual elections results deviate from a theoretically or empirically determined \emph{seats-votes curve}. Such function, first introduced into political science by \citet{Butler50} with his rediscovery of the \emph{cube law}, have been intensely studied from the 1950-s to the 1990-s (see, e.g., \citealp{KendallStuart50,Brookes53,March57,Theil69,Taagepera73,Tufte73,LinehanSchrodt77,Grofman83,BrowningKing87,CampagnaGrofman90,BradyGrofman91,GarandParent91,GilliganMatsusaka99}). There is a somewhat broad consensus in the literature that a two-party seats-votes relation is usually described by a modified power law:
\begin{equation}
    \frac{s_i}{1-s_i} = \beta_i \left(\frac{v_i}{1-v_i}\right)^{\rho},
\end{equation}
where $s_i$ and $v_i$ are, respectively, the seat- and vote-share of the $i$-th party, $\beta_i$ is a party-dependent parameter, and $\rho$ is a constant \citep{Tufte73,Grofman83}. However, only few authors have considered the case of multiparty elections \citep{Taagepera86,King90,Linzer12}, and their results are mostly heuristic in nature, lacking formal theoretical grounding.

The state-of-the-art approach to detecting gerrymandering is the \emph{partisan symmetry} method. The general concept was first proposed by \citet{NiemiDeegan78}, who noted that an election should not be regarded as gerrymandered if it deviates from a model seats-votes curve as long as the deviation is the same for each party, i.e., each party has the same seats-votes curve. The main challenge here lies in obtaining that curve from a single realization. The original idea has been to extrapolate therefrom by assuming a \emph{uniform partisan swing}, i.e., that as the aggregate vote share of a party changes, its district-level vote shares increase or decrease uniformly and independently of their original levels. This assumption, first proposed by \citet{Butler47,Butler51}, has been employed by, inter alia, \citet{SoperRydon58}, \citet{Brookes59,Brookes60}, \citet{Tufte73}, \citet{BackstromEtAl78}, \citet{GudginTaylor79}, \citet{Scarrow81,Scarrow82}, \citet{Niemi85}, \citet{NiemiFett86}, \citet{GarandParent91}, \citet{Aistrup95}, \citet{JohnstonEtAl99}, and others. However, in light of both theoretical and empirical criticism of the uniform partisan swing assumption \citep{McLean73,Basehart87,Jackman94,Blau01}, a more sophisticated extrapolation method has been developed by \citet{GelmanKing90,GelmanKing90a,GelmanKing94}, see also \citealp{King89,KingGelman91}, and \citealp{ThomasEtAl13}. However, neither of these two methods can account for multiple parties absent some unrealistic assumption that the relevant swing happens only between two parties identified in advance as major, but see an attempt to develop a multi-party variant of the Gelman-King method by \citealt{Monroe98}.

The third approach is the \emph{efficiency gap} method proposed by \citet{McGhee14} and further developed in \citet{StephanopoulosMcGhee15}. It is based on the assumption that in an unbiased election all contending parties should waste the same number of votes. While \textit{prima facie} attractive, this assumption is actually highly problematic because it requires the electoral system to match a very specific seats-votes curve [\citealp[p.~296]{McGannEtAl15}, \citealp{BernsteinDuchin17}]. In this respect it represents a methodological step backwards, making it again impossible to distinguish asymmetry from responsiveness. The McGhee-Stephanopoulos definition of wasted votes has also been criticized as counterintuitive [see, e.g., \citealp[pp.~1181-84]{Cover18}, and \citealp[p.~5]{BestEtAl18}]. From our perspective the primary weakness of the efficiency gap method, as well as of its many variants \citep{Nagle17,Cover18,Dopp18,Veomett18,Tapp18,Leibzon22} is again the lack of accounting for multiple parties. The original efficiency gap criterion is violated in almost every multiparty election.


Finally, there are several method designed to identify anomalies in the vote distribution indicative of standard gerrymandering techniques like packing and cracking. These include the \emph{mean-median difference} test proposed by \citet{McDonaldEtAl11}, which measures the skewness of the vote distribution; the \emph{multimodality} test put forward by \citet{Erikson72}; the \emph{declination coefficient} introduced by \citet{Warrington18} and measuring the change in the shape of the cumulative distribution function of vote shares at $1/2$; and the \emph{lopsided winds} method of testing whether the difference between the winners' vote shares in districts won by the first and the second party is statistically significant \citep{Wang16}. Again, virtually of all those methods assume a two-party system. For instance, natural marginal vote share distributions in multiparty systems (such as the beta distribution or the log-normal distribution) are necessarily skewed. A similar assumptions underlies the declination ratio and the lopsided wins test. The multimodality test, on the other hand, assumes a constant number of competitors.

\subsection{Basic Concepts and Notation}

Gerrymandering is usually defined as manipulation of electoral district boundaries aimed at achieving a political benefit. Hence, \emph{intentionality} is inherent in the very concept. However, identical results can also arise non-intentionally. For instance, geographic concentration of one party's electorate in small areas (major cities, regions) can produce similar effects to intentional packing. We use the term `electoral bias' to refer to such `nonintentional gerrymandering'.

Our basic idea is to treat gerrymandering and electoral bias as \emph{statistical anomalies} in the \emph{translation of votes into seats}. Identification of such anomalies requires a reference point, either theoretical, such as a theoretical model of district-level vote distribution, or empirical, such as a large set of other elections that can be expected to have come from the same statistical population. The former approach is undoubtedly more elegant, but burdened with the risk that the theoretical model deviates from the empirical reality. Hence, in this paper we focus on the empirical approach.

There are three basic assumptions underlying our methodology. One is that we have a \emph{training set} of elections that come from the same statistical population as the election we are studying. Another one is that gerrymandering (or any other form of electoral bias) is \emph{an exception rather than a rule}. Thus, we assume that a substantial majority of the training set elections are free from bias. The third assumption is that while district-level results of individual candidates can be tainted by gerrymandering, \emph{aggregate electoral results} (e.g., vote shares) never are.

One major limitation of our methodology lies in its \emph{inability to distinguish gerrymandering from natural electoral bias}. This limitation is shared, however, with virtually all methods in which the evidence for gerrymandering is sought in analyzing voting patterns or any other variables which are ultimately a function of such patterns (e.g., seat shares, wasted votes, etc.). For many applications that may be enough, since for many potential end-users of our method it might not matter whether the bias in the electoral system is artificial or natural. Even for applications where that distinction matters, the proposed methods might still be useful to identify cases requiring more in-depth investigation, which is usually necessary to find evidence of the intent to gerrymander.

Let us introduce some basic notation to be used throughout this paper:
\begin{description}
    \item[set of districts] We denote the set of districts by $D := \{1, \dots, c\}$.
    \item[set of parties] We denote the set of parties by $P := \{1, \dots, n\}$.
    \item[set of contested districts] For $i \in P$, we denote the set of districts in which the $i$-th party runs a candidate by $D_i$. Let $c_i := |P_i|$.
    \item[set of contesting parties] For $k \in D$, we denote the set of parties that run a candidate in the $k$-th district by $P_k$. Let $n_k := |D_k|$.
    \item[district-level vote share] For $i \in P$ and $k \in D$, we denote the district-level vote share of the $i$-th party's candidate in the $k$-th district by $v_i^k$. If there was no such candidate, we assume $v_i^k = 0$.
    \item[district-level seat share] For $i \in P$ and $k \in D$, let $s_i^k$ equal $1$ if the $i$-th party's candidate in the $k$-th district won the seat, and $0$ otherwise.
    \item[district size] For $k \in D$, we denote the number of voters cast in the $k$-th district by $w_k$.
    \item[aggregate vote share] For $i \in P$, we denote the aggregate vote share of the $i$-th party by $v_i := \left(\sum_{k \in D_i} v_i^k w_k\right) / \left(\sum_{k \in D_i} w_k\right).$
    \item[aggregate seat share] For $i \in P$, we denote the aggregate seat share of the $i$-th party by $s_i := \left(\sum_{k \in D_i} s_i^k\right) / c_i.$
    \item [unit simplex] For $n \in \mathbb{N}_{+}$, we denote the $(n-1)$-dimensional unit simplex by $\Delta_n := \{\mathbf{x} \in \mathbb{R}_{+}^{n} : \lVert\mathbf{x}\rVert_{1} = 1\}$.
    \item [$k$-th largest / smallest coordinate] For $n \in \mathbb{N}_{+}$, $\mathbf{x} \in \mathbb{R}^{n}$, and $k \in 1, \dots, n$, we denote the $(k)$-th largest coordinate of $\mathbf{x}$ by $x_k^{\downarrow}$, and the $(k)$-the smallest coordinate of $\mathbf{x}$ by $x_k^{\uparrow}$.
\end{description}

\begin{definition}[Uniform Distribution]
    For any $n \in \mathbb{N}_{+}$, a \emph{Uniform distribution} on the unit simplex $\Delta_n$, denoted by $\Unif(\Delta_n)$, is the unique absolutely continuous probability distribution supported on $\Delta_n$ whose density with respect to the Lebesgue measure on $\Delta_n$ is constant.
\end{definition}

\begin{definition}[Dirichlet Distribution]
    For any $n \in \mathbb{N}_{+}$, a \emph{Dirichlet distribution} on the unit simplex $\Delta_n$ is any absolutely continuous probability distribution supported on $\Delta_n$ and parametrized by a vector $\boldsymbol{\alpha} \in \mathbb{R}_{+}^n$ whose density with respect to the Lebesgue measure on $\Delta_n$ is given by:
    \begin{equation}
    f(\mathbf{x}) = \frac{1}{B(\mathbf{\alpha})} \prod_{i=1}^{n} x_i^{\alpha_i - 1}.
    \end{equation}
\end{definition}

\begin{remark}
    Note that $\Dir(\{1\}^{n}) = \Unif(\Delta_n)$ for any $n \in \mathbb{N}_{+}$.
\end{remark}

\section{Seats-Votes Functions}\label{sec:sv}

\emph{Seats-votes curves} are one of the fundamental concepts under the traditional approach to the quantitative study of electoral systems. It is a function that maps an aggregate vote share to an aggregate seat share. Of course, it is easy to see that in reality even in two-party elections a seats-votes curve is not actually real-valued, but probability measure-valued, since the seat share depends on what we call `electoral geography' -- the distribution of district-level vote shares. We call this measure-valued function a \emph{seats-votes function}, while reserving the name of a seats-votes curve to a function that maps a vote share to the expectation of its image under the seats-votes function. Note that both gerrymandering and electoral bias manifest themselves by deviation of the seats-votes function applicable to one or more parties from the `model' seats-votes function (however the latter is determined) caused by anomalies of the electoral geography.

In multi-party elections there is another fundamental problem with seats-votes functions: the distribution of seats depends not only on the vote share and the electoral geography, but also on the competition patterns: the number of competitors and the distribution of their votes (or, to be more precise, on the distribution of the first order statistic of their votes) \citep{Calvo09,Manow11,CalvoRodden15}. If we were to fit a single seats-votes for all parties without regard to competition patterns, the result would involve another source of randomness besides districting effects, namely the variation in such patterns. Hence, we would be unable to distinguish between a seats-votes function that deviates from the model because of electoral geography and a seats-votes function that also deviates from the model, but because of unusual competition patterns. Thus, we need to account for this effect by considering a seats-votes-competition pattern function rather than the usual seats-votes function.

\begin{remark}
    Consider seats-votes curves in multi-party elections. If we assume that they are anonymous (i.e., identical for all parties), non-decreasing, and surjective, it turns out perfect proportionality ($s = v$) is the only seats-votes curve that does not depend on the distribution of competitors' votes \citep[Theorem~1]{BoratynEtAl22}.
\end{remark}

It would be convenient if we were able to describe the competition pattern by a single numerical parameter. Our objective here is to find a measure of the `difficulty' of winning a seat given the number of competitors and the distribution of their vote shares (renormalized so as to sum to $1$). A natural choice would be the \emph{seat threshold}:

\begin{definition}[Seat Threshold]
    Fix $i = 1, \dots, n$, and assume that renormalized vote shares of the competitors of the $i$-th candidate equal some random variable $\mathbf{Z}$ distributed according to some probability measure on $\Delta_{n,-i}$. A \emph{seat threshold} of the $i$-th candidate is such $t_i \in [1/n, 1/2]$ that $Pr(S_i = 1 | v_i) > 1/2$ for every $v_i > t_i$, i.e., the probability that the $i$-th candidate wins a seat with vote share equal $v_i$ exceeds $1/2$.
\end{definition}

\begin{proposition}
    It is easy to see that $Pr(S_i = 1 | v) = 1 - F_{Z_{1}^{\downarrow}}(v / (1-v))$, where $F_{Z_{1}^{\downarrow}}$ is the cumulative distribution function of the renormalized vote share of the largest competitor.
\end{proposition}

Out next objective is to approximate the seat threshold in cases where we do not have any knowledge of the distribution of the competitors' vote shares, but only a single realization thereof. We therefore need a statistic that is both a stable estimator of the distribution parameters and highly correlated with the value of the largest order statistic. We posit that the best candidates for such statistics are \emph{measures of vote diversity} among competitors, and use a Monte Carlo simulation to test a number of such measures commonly considered in social sciences.

\begin{observation}
    Let $\alpha \sim \mathrm{Gamma}(1,1)$, and let $\mathbf{V} \sim \Dir(\{\alpha\}^n)$, where $n = 3, \dots, 12$. On a sample of $2^{16}$ independent realizations of $\mathbf{V}$ we have calculated Spearman's rank correlation coefficients \cite{Spearman04} of the following variables:
    \begin{enumerate}
        \item $\alpha$,\smallskip
        \item $\mathbf{V}_1^{\downarrow}$, i.e., maximum of the coordinates,\smallskip
        \item $\mathbf{V}_1^{\uparrow}$, i.e., minimum of the coordinates,\smallskip
        \item median coordinate $V_{\mathrm{med}}$,\smallskip
        \item \emph{Shannon entropy} \cite{Shannon48}, $\mathrm{H}(\mathbf{V}) := -\sum_{i=1}^n V_i \log V_i$,\smallskip
        \item \emph{Herfindahl--Hirschman--Simpson index} \cite{Hirschman45,Simpson49,Herfindahl50}, $\sum_{i=1}^n V_i^2$,
        \item \emph{Gini coefficient} of the coordinates,
        \item \emph{Bhattacharyya angle} \cite{Bhattacharyya43} between $V$ and the barycenter of the simplex, $\arccos \sum_{i=1}^n \sqrt{V_i/n}$.
    \end{enumerate}\medskip

    The results for $n = 3, 6, 12$ are given in, respectively, \autoref{tbl:divers3}, \autoref{tbl:divers6}, and \autoref{tbl:divers12}.

    \begin{table}[htb]
    \begin{tabular}{lcccccccc}
    \toprule
    \textbf{} & \textbf{$\alpha$} & \textbf{$V_1^{\downarrow}$} & \textbf{$V_1^{\uparrow}$} & \textbf{$V_{\mathrm{med}}$} & \textbf{$\mathrm{H}(\mathbf{V})$} & \textbf{$\sum_{i=1}^n V_i^2$} & \textbf{Gini} & \textbf{Bhatt.} \\
    \midrule
    \textbf{$\alpha$} & \cellcolor[HTML]{63BE7B}1.000 & \cellcolor[HTML]{FBAF78}-0.513 & \cellcolor[HTML]{9ECF7F}0.583 & \cellcolor[HTML]{CEDD82}0.246 & \cellcolor[HTML]{9FD07F}0.582 & \cellcolor[HTML]{FBA877}-0.565 & \cellcolor[HTML]{FBA777}-0.568 & \cellcolor[HTML]{FBA476}-0.588 \\
    \textbf{$V_1^{\downarrow}$} & \cellcolor[HTML]{FBAF78}-0.513 & \cellcolor[HTML]{63BE7B}1.000 & \cellcolor[HTML]{F98470}-0.806 & \cellcolor[HTML]{FA9072}-0.729 & \cellcolor[HTML]{F8716C}-0.938 & \cellcolor[HTML]{68C07C}0.965 & \cellcolor[HTML]{68C07C}0.965 & \cellcolor[HTML]{6FC27C}0.917 \\
    \textbf{$V_1^{\uparrow}$} & \cellcolor[HTML]{9ECF7F}0.583 & \cellcolor[HTML]{F98470}-0.806 & \cellcolor[HTML]{63BE7B}1.000 & \cellcolor[HTML]{D1DE82}0.226 & \cellcolor[HTML]{6AC07C}0.952 & \cellcolor[HTML]{F8746D}-0.918 & \cellcolor[HTML]{F8726C}-0.930 & \cellcolor[HTML]{F86D6B}-0.967 \\
    \textbf{$V_{\mathrm{med}}$} & \cellcolor[HTML]{CEDD82}0.246 & \cellcolor[HTML]{FA9072}-0.729 & \cellcolor[HTML]{D1DE82}0.226 & \cellcolor[HTML]{63BE7B}1.000 & \cellcolor[HTML]{ACD380}0.485 & \cellcolor[HTML]{FBA877}-0.564 & \cellcolor[HTML]{FBAA77}-0.549 & \cellcolor[HTML]{FCBA7A}-0.436 \\
    \textbf{$\mathrm{H}(\mathbf{V})$} & \cellcolor[HTML]{9FD07F}0.582 & \cellcolor[HTML]{F8716C}-0.938 & \cellcolor[HTML]{6AC07C}0.952 & \cellcolor[HTML]{ACD380}0.485 & \cellcolor[HTML]{63BE7B}1.000 & \cellcolor[HTML]{F8696B}-0.994 & \cellcolor[HTML]{F8696B}-0.993 & \cellcolor[HTML]{F8696B}-0.998 \\
    \textbf{$\sum_{i=1}^n   V_i^2$} & \cellcolor[HTML]{FBA877}-0.565 & \cellcolor[HTML]{68C07C}0.965 & \cellcolor[HTML]{F8746D}-0.918 & \cellcolor[HTML]{FBA877}-0.564 & \cellcolor[HTML]{F8696B}-0.994 & \cellcolor[HTML]{63BE7B}1.000 & \cellcolor[HTML]{64BF7C}0.997 & \cellcolor[HTML]{65BF7C}0.986 \\
    \textbf{Gini} & \cellcolor[HTML]{FBA777}-0.568 & \cellcolor[HTML]{68C07C}0.965 & \cellcolor[HTML]{F8726C}-0.930 & \cellcolor[HTML]{FBAA77}-0.549 & \cellcolor[HTML]{F8696B}-0.993 & \cellcolor[HTML]{64BF7C}0.997 & \cellcolor[HTML]{63BE7B}1.000 & \cellcolor[HTML]{65BF7C}0.986 \\
    \textbf{Bhatt.} & \cellcolor[HTML]{FBA476}-0.588 & \cellcolor[HTML]{6FC27C}0.917 & \cellcolor[HTML]{F86D6B}-0.967 & \cellcolor[HTML]{FCBA7A}-0.436 & \cellcolor[HTML]{F8696B}-0.998 & \cellcolor[HTML]{65BF7C}0.986 & \cellcolor[HTML]{65BF7C}0.986 & \cellcolor[HTML]{63BE7B}1.000 \\
    \bottomrule
    \end{tabular}
    \label{tbl:divers3}
    \caption{Correlation Matrix for $n = 3$}
    \end{table}

    \begin{table}[htb]
    \begin{tabular}{lllllllll}
    \toprule
    \textbf{} & \textbf{$\alpha$} & \textbf{$V_1^{\downarrow}$} & \textbf{$V_1^{\uparrow}$} & \textbf{$V_{\mathrm{med}}$} & \textbf{$\mathrm{H}(\mathbf{V})$} & \textbf{$\sum_{i=1}^n V_i^2$} & \textbf{Gini} & \textbf{Bhatt.} \\
    \midrule
    \textbf{$\alpha$} & \cellcolor[HTML]{63BE7B}1.000 & \cellcolor[HTML]{FA9D75}-0.643 & \cellcolor[HTML]{89C97E}0.728 & \cellcolor[HTML]{AED480}0.462 & \cellcolor[HTML]{86C87D}0.751 & \cellcolor[HTML]{FA9172}-0.724 & \cellcolor[HTML]{FA8F72}-0.739 & \cellcolor[HTML]{F98B71}-0.762 \\
    \textbf{$V_1^{\downarrow}$} & \cellcolor[HTML]{FA9D75}-0.643 & \cellcolor[HTML]{63BE7B}1.000 & \cellcolor[HTML]{FA9773}-0.683 & \cellcolor[HTML]{F9826F}-0.825 & \cellcolor[HTML]{F8756D}-0.910 & \cellcolor[HTML]{6AC07C}0.952 & \cellcolor[HTML]{6EC17C}0.925 & \cellcolor[HTML]{75C37C}0.876 \\
    \textbf{$V_1^{\uparrow}$} & \cellcolor[HTML]{89C97E}0.728 & \cellcolor[HTML]{FA9773}-0.683 & \cellcolor[HTML]{63BE7B}1.000 & \cellcolor[HTML]{B9D780}0.386 & \cellcolor[HTML]{74C37C}0.882 & \cellcolor[HTML]{F98370}-0.820 & \cellcolor[HTML]{F97D6F}-0.856 & \cellcolor[HTML]{F8746D}-0.918 \\
    \textbf{$V_{\mathrm{med}}$} & \cellcolor[HTML]{AED480}0.462 & \cellcolor[HTML]{F9826F}-0.825 & \cellcolor[HTML]{B9D780}0.386 & \cellcolor[HTML]{63BE7B}1.000 & \cellcolor[HTML]{8FCB7E}0.688 & \cellcolor[HTML]{F98C71}-0.756 & \cellcolor[HTML]{FA9172}-0.721 & \cellcolor[HTML]{FA9E75}-0.636 \\
    \textbf{$\mathrm{H}(\mathbf{V})$} & \cellcolor[HTML]{86C87D}0.751 & \cellcolor[HTML]{F8756D}-0.910 & \cellcolor[HTML]{74C37C}0.882 & \cellcolor[HTML]{8FCB7E}0.688 & \cellcolor[HTML]{63BE7B}1.000 & \cellcolor[HTML]{F8696B}-0.990 & \cellcolor[HTML]{F8696B}-0.995 & \cellcolor[HTML]{F8696B}-0.995 \\
    \textbf{$\sum_{i=1}^n   V_i^2$} & \cellcolor[HTML]{FA9172}-0.724 & \cellcolor[HTML]{6AC07C}0.952 & \cellcolor[HTML]{F98370}-0.820 & \cellcolor[HTML]{F98C71}-0.756 & \cellcolor[HTML]{F8696B}-0.990 & \cellcolor[HTML]{63BE7B}1.000 & \cellcolor[HTML]{64BF7C}0.993 & \cellcolor[HTML]{67C07C}0.974 \\
    \textbf{Gini} & \cellcolor[HTML]{FA8F72}-0.739 & \cellcolor[HTML]{6EC17C}0.925 & \cellcolor[HTML]{F97D6F}-0.856 & \cellcolor[HTML]{FA9172}-0.721 & \cellcolor[HTML]{F8696B}-0.995 & \cellcolor[HTML]{64BF7C}0.993 & \cellcolor[HTML]{63BE7B}1.000 & \cellcolor[HTML]{66BF7C}0.985 \\
    \textbf{Bhatt.} & \cellcolor[HTML]{F98B71}-0.762 & \cellcolor[HTML]{75C37C}0.876 & \cellcolor[HTML]{F8746D}-0.918 & \cellcolor[HTML]{FA9E75}-0.636 & \cellcolor[HTML]{F8696B}-0.995 & \cellcolor[HTML]{67C07C}0.974 & \cellcolor[HTML]{66BF7C}0.985 & \cellcolor[HTML]{63BE7B}1.000 \\
    \bottomrule
    \end{tabular}
    \label{tbl:divers6}
    \caption{Correlation Matrix for $n = 6$}
    \end{table}

    \begin{table}[]
    \begin{tabular}{lllllllll}
    \toprule
    \textbf{} & \textbf{$\alpha$} & \textbf{$V_1^{\downarrow}$} & \textbf{$V_1^{\uparrow}$} & \textbf{$V_{\mathrm{med}}$} & \textbf{$\mathrm{H}(\mathbf{V})$} & \textbf{$\sum_{i=1}^n V_i^2$} & \textbf{Gini} & \textbf{Bhatt.} \\
    \midrule
    \textbf{$\alpha$} & \cellcolor[HTML]{63BE7B}1.000 & \cellcolor[HTML]{FA8E72}-0.730 & \cellcolor[HTML]{7EC67D}0.820 & \cellcolor[HTML]{9CCF7F}0.612 & \cellcolor[HTML]{78C47D}0.856 & \cellcolor[HTML]{F9806F}-0.829 & \cellcolor[HTML]{F97D6F}-0.850 & \cellcolor[HTML]{F87B6E}-0.868 \\
    \textbf{$V_1^{\downarrow}$} & \cellcolor[HTML]{FA8E72}-0.730 & \cellcolor[HTML]{63BE7B}1.000 & \cellcolor[HTML]{FA9473}-0.690 & \cellcolor[HTML]{F98670}-0.787 & \cellcolor[HTML]{F8766D}-0.902 & \cellcolor[HTML]{6CC17C}0.944 & \cellcolor[HTML]{72C37C}0.900 & \cellcolor[HTML]{76C47D}0.871 \\
    \textbf{$V_1^{\uparrow}$} & \cellcolor[HTML]{7EC67D}0.820 & \cellcolor[HTML]{FA9473}-0.690 & \cellcolor[HTML]{63BE7B}1.000 & \cellcolor[HTML]{A7D27F}0.532 & \cellcolor[HTML]{77C47D}0.868 & \cellcolor[HTML]{F9826F}-0.815 & \cellcolor[HTML]{F97D6E}-0.853 & \cellcolor[HTML]{F8766D}-0.902 \\
    \textbf{$V_{\mathrm{med}}$} & \cellcolor[HTML]{9CCF7F}0.612 & \cellcolor[HTML]{F98670}-0.787 & \cellcolor[HTML]{A7D27F}0.532 & \cellcolor[HTML]{63BE7B}1.000 & \cellcolor[HTML]{84C87D}0.772 & \cellcolor[HTML]{F98370}-0.810 & \cellcolor[HTML]{F98670}-0.788 & \cellcolor[HTML]{F98D72}-0.737 \\
    \textbf{$\mathrm{H}(\mathbf{V})$} & \cellcolor[HTML]{78C47D}0.856 & \cellcolor[HTML]{F8766D}-0.902 & \cellcolor[HTML]{77C47D}0.868 & \cellcolor[HTML]{84C87D}0.772 & \cellcolor[HTML]{63BE7B}1.000 & \cellcolor[HTML]{F8696B}-0.991 & \cellcolor[HTML]{F8696B}-0.998 & \cellcolor[HTML]{F8696B}-0.996 \\
    \textbf{$\sum_{i=1}^n   V_i^2$} & \cellcolor[HTML]{F9806F}-0.829 & \cellcolor[HTML]{6CC17C}0.944 & \cellcolor[HTML]{F9826F}-0.815 & \cellcolor[HTML]{F98370}-0.810 & \cellcolor[HTML]{F8696B}-0.991 & \cellcolor[HTML]{63BE7B}1.000 & \cellcolor[HTML]{65BF7C}0.990 & \cellcolor[HTML]{67C07C}0.976 \\
    \textbf{Gini} & \cellcolor[HTML]{F97D6F}-0.850 & \cellcolor[HTML]{72C37C}0.900 & \cellcolor[HTML]{F97D6E}-0.853 & \cellcolor[HTML]{F98670}-0.788 & \cellcolor[HTML]{F8696B}-0.998 & \cellcolor[HTML]{65BF7C}0.990 & \cellcolor[HTML]{63BE7B}1.000 & \cellcolor[HTML]{65BF7C}0.992 \\
    \textbf{Bhatt.} & \cellcolor[HTML]{F87B6E}-0.868 & \cellcolor[HTML]{76C47D}0.871 & \cellcolor[HTML]{F8766D}-0.902 & \cellcolor[HTML]{F98D72}-0.737 & \cellcolor[HTML]{F8696B}-0.996 & \cellcolor[HTML]{67C07C}0.976 & \cellcolor[HTML]{65BF7C}0.992 & \cellcolor[HTML]{63BE7B}1.000 \\
    \bottomrule
    \end{tabular}
    \label{tbl:divers12}
    \caption{Correlation Matrix for $n = 12$}
    \end{table}
\end{observation}

    We conclude that the \emph{Herfindahl--Hirschman--Simpson index} is consistently the one that best correlates with the maximal coordinate while also being a reasonably good estimate of the distribution parameters. Accordingly, in our procedure for estimating the seat threshold we use its monotonic transform, i.e., the \emph{effective number of competitors} \cite{LaaksoTaagepera79,TaageperaGrofman81}:

\begin{definition}[Effective Number of Competitors]
    The \emph{effective number of competitors} of the $i$-th candidate, $i = 1, \dots, n$, is given by:
    \begin{equation}
        \varphi := \left(\sum_{j=1, j \neq i}^{n} z_j^2\right)^{-1},
    \end{equation}
    where $\mathbf{z} \in \Delta_{n,-i}$ is a vector of the vote shares of that candidate's competitors multiplied by such constant in $\mathbb{R}_{+}$ that $\sum_{j=1, j \neq i}^{n} z_j = 1$.
\end{definition}

\begin{figure}[htb]
    \centering
    \includegraphics[width=0.49\textwidth]{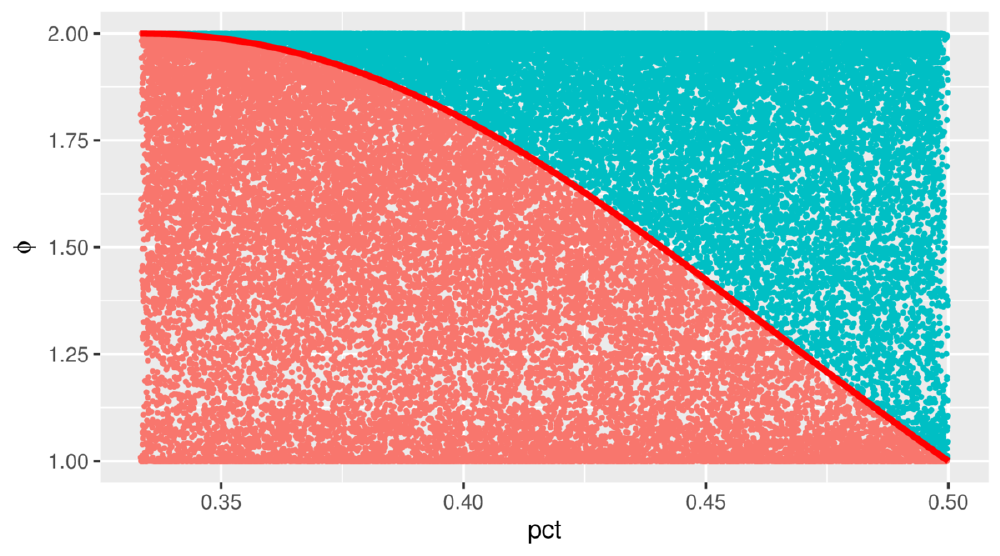}
    \includegraphics[width=0.49\textwidth]{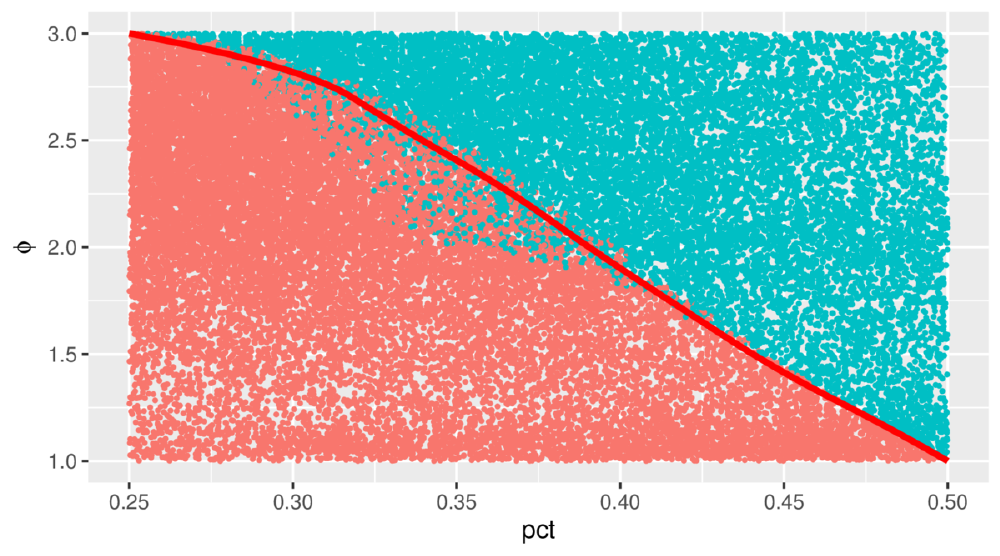}
    \includegraphics[width=0.49\textwidth]{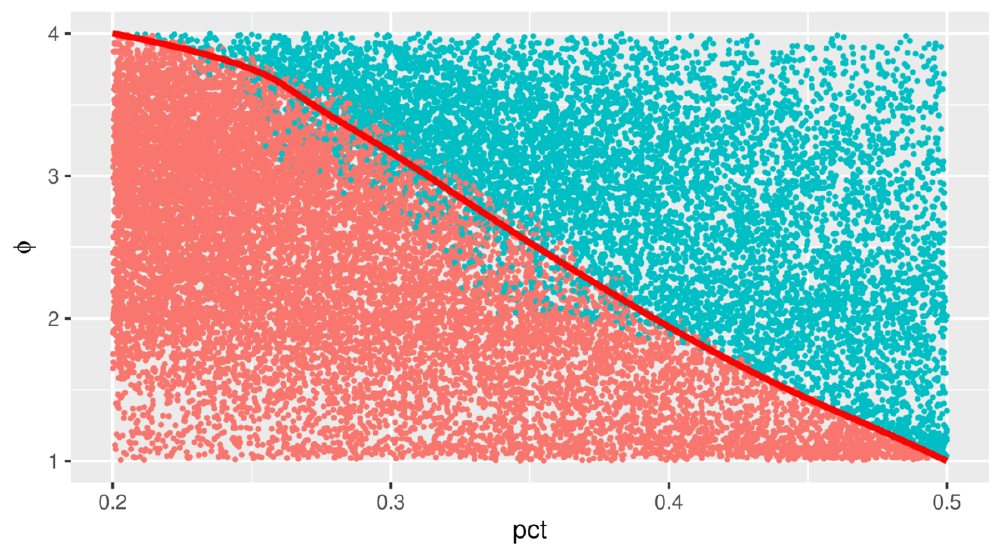}
    \includegraphics[width=0.49\textwidth]{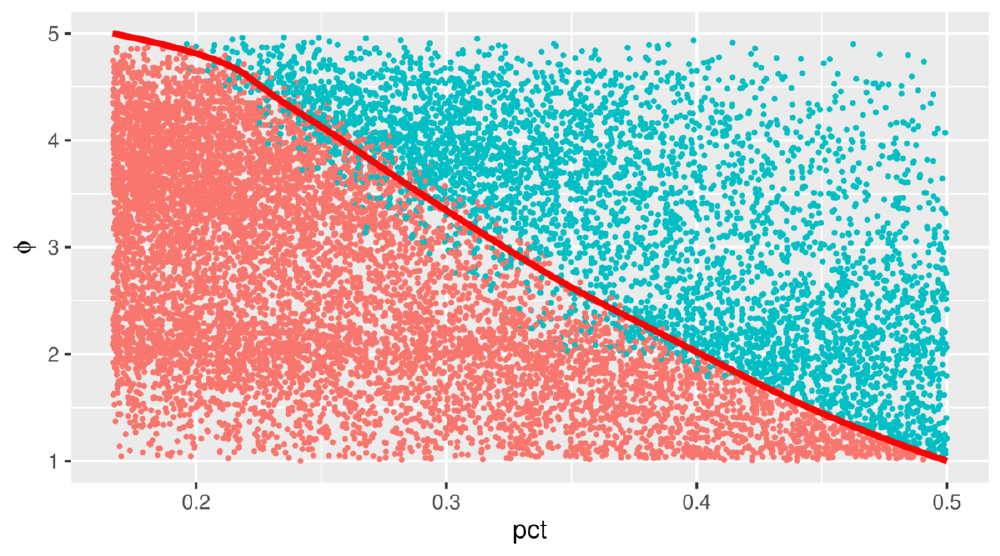}
    \caption{Effective seat thresholds for $n = 3, 4, 5, 6$. Blue points indicate successful candidates, while red points -- unsuccessful candidates.}
    \label{fig:estClassif}
\end{figure}

    We shall see that vote share, the number of competitors, and the effective number of competitors enable us to classify candidates as winning and losing with a quite small classification error (see \autoref{fig:estClassif} and \autoref{tbl:estClassifErr}).
    
\begin{proposition}
    Clearly, with three candidates, i.e., two competitors, the classifier is exact (modulo ties), as the effective number of competitors uniquely determines the share of the larger one in their aggregate vote share:
    \begin{equation}
        \max \{z_{j_1}, z_{j_2}\} = \frac{1}{2} \left(1 + \sqrt{\frac{2}{\varphi_i} - 1}\right).
    \end{equation}
    Then the decision boundary (i.e., the curve separating the space of candidates into winning and losing subspaces) is the set of points satisfying:
    \begin{equation}
        \varphi_i = \frac{1 - 2 v_i + v_i^2}{1 - 4 v_i + 5 v_i^2}.
    \end{equation}
\end{proposition}

\begin{model}[Decision Boundary for $n > 3$]
    For $n > 3$, the decision boundary is determined on the basis of the data using a support vector machine-based classifier \cite{BoserEtAl92,CortesVapnik95} with a third-order polynomial kernel, and then smoothed by approximating it with a strictly decreasing B-spline of degree $3$, with boundary nodes at $1/n$ and $1/2$ and interior nodes fitted using cross-validation.
\end{model}

\begin{definition}[Effective Seat Threshold]
    We refer to the value of the decision boundary for the candidate of the $i$-th party in the $k$-th district, ascertained for the empirically given number and effective number of competitors, as the \emph{effective seat threshold} of the $i$-th party in the $k$-th district, and denote it by $t_i^k$.
\end{definition}

\begin{definition}[Effective Seat Threshold Classifier]
    An effective seat threshold classifier is a function $\mathfrak{s} : [0, 1] \times \mathbb{N} \times [1, \infty) \rightarrow \mathbb{B}$ that maps a triple $(v, n, \varphi)$ to $0$ if the probability of winning a seat with vote share $v$, $n-1$ competitors, and $\varphi$ effective competitors is below $1/2$, and to $1$ otherwise.
\end{definition}

\begin{table}[htb]
    \begin{tabular}{cccccccc}
    \toprule
    $\boldsymbol{n}$ & $\boldsymbol{R}$ & \hspace{2.5em} & $\boldsymbol{n}$ & $\boldsymbol{R}$ & \hspace{2.5em} & $\boldsymbol{n}$ & $\boldsymbol{R}$ \\
    \midrule
    3 & .0035 & & 9 & .0073 & & 15 & .0144 \\
    4 & .0137 & & 10 & .0067 & & 16 & .0148 \\
    5 & .0152 & & 11 & .0142 & & 17 & .0176 \\
    6 & .0137 & & 12 & .0186 & & 18 & .0182 \\
    7 & .0136 & & 13 & .0152 & & 19 & .0133 \\
    8 & .0068 & & 14 & .0171 & & 20 & .0168 \\
    \bottomrule
    \end{tabular}
    \label{tbl:estClassifErr}
    \caption{Classification error $R$ for the effective seat threshold classifier.}
\end{table}

\begin{definition}[Mean Effective Seat Threshold]
    \emph{Mean effective seat threshold}, $t_i := \langle t_i^k \rangle_{k \in D_i}$, where $D_i$ is the set of districts contested by the $i$-th party, is our measure of the difficulty of winning a seat.
\end{definition}

\section{Nonparametric Seats-Votes Function Estimates}\label{sec:np}

As already noted in Sec. 1, one possible approach to identifying the model seats-votes function is to construct one theoretically. We might start with some probabilistic model of intra-district vote distribution, then use it to calculate the seat threshold, and finally use a probabilistic model of inter-district vote distribution to calculate the probability of district vote share exceeding the seat threshold. Finally, either by convolving binomial distributions (for small values of $c$) or by the central limit theorem (for large values of $c$) we obtain the expected seat share, as well as the distribution around the mean.

One unavoidable weakness of any theoretical seats-votes curve lies in the fact that a systematic deviation therefrom might just as easily arise from gerrymandering or any other electoral bias as from incongruities between the theoretical distributional assumptions and the empirical reality. To avoid this issue we derive our model seats-votes function solely from the reference election dataset with minimal theoretical assumptions\footnote{In particular, we assume that the seat shares are distributed according to some absolutely continuous probability measure supported on $[0, 1]$} by using the kernel regression method \citep{Nadaraya65,Watson64} to obtain an estimate of the conditional cumulative distribution function of a seat share for the given vote share. Its general idea is to estimate the conditional expectation of a random variable at a point in the condition space by averaging the values of its realizations at neighboring points with distance-decreasing weights. Because the method can be sensitive as to the choice of that method's hyperparameters, we discuss those choices in some detail.

\begin{model}[Locally-Constant Kernel Regression]\label{mod:npreg}
    Let $S \in \mathbb{R}$ be a random \emph{response variable}, and let $\mathbf{X} \in \mathfrak{F}$, where $\mathfrak{F}$ is some linear \emph{feature space} and $D := \dim \mathfrak{F}$, be a vector of \emph{predictor variables}. Assume we have a vector of $N$ realizations of $S$, $\mathbf{s}$, and an $N \times \dim \mathfrak{F}$ matrix of realizations of $\mathbf{X}$, $\mathbf{x}$. We denote its $j$-th row by $\mathbf{x}_{j}$. Then the \emph{locally-linear kernel regression estimate} of the conditional expectation of $S$ given a vector of predictors $\mathbf{x}_{0} \in \mathfrak{F}$ is given by:
    \begin{equation}
        \mathbb{E}(S | \mathbf{x}_{0}) = \frac{\sum_{j=1}^{N} s_j K\left((\mathbf{x}_j - \mathbf{x}_{0}) \mathbf{h}_{\mathbf{x}_{0}, \mathbf{x}_j}\right)}{\sum_{j=1}^{N} K\left((\mathbf{x}_j - \mathbf{x}_{0}) \mathbf{h}_{\mathbf{x}_{0}, \mathbf{x}_j}\right)},
    \end{equation}
    where $N$ is the number of observations (in our case -- sum of the number of parties over all elections in our set of elections), $K$ is a second-order \emph{kernel}, and $\mathbf{h}_{\mathbf{x}_{0}, \mathbf{x}_j} \in \mathbb{R}_{+}^{D}$ is a \emph{bandwidth} parameter for the pair $(\mathbf{x}_{0}, \mathbf{x}_j)$. In other words, we average the values of $s$ over all parties with weights determined by the value of the kernel at $(\mathbf{x}_j - \mathbf{x}_{0}) \mathbf{h}_{\mathbf{x}_{0}, \mathbf{x}_j}$.
\end{model}

\noindent\textbf{Choice of Kernel.}

\begin{definition}[Kernel]
    A $k$-order \emph{kernel}, where $k \ge 2 \in \mathbb{N}$, is any function $\mathbb{R}^{D} \rightarrow [0, \infty)$ of class $C^{k}$ satisfying:
    \begin{itemize}
        \item $\int_{\mathbb{R}^{D}} K(\mathbf{x}) d\mathbf{x} = 1$,
        \item $\int_{\mathbb{R}^{D}} \mathbf{x} K(\mathbf{x}) d\mathbf{x} = 0$,
        \item $\int_{\mathbb{R}^{D}} \mathbf{x}^k K(\mathbf{x}) d\mathbf{x} \in \mathbb{R}^{D}$.
    \end{itemize}
\end{definition}

\noindent \textit{Prima facie} it would seem that the appropriate choice of the kernel is fundamental in fitting a kernel density model. However, this is not actually the case: most of the commonly used kernels, including Gaussian, Epanechnikov (square), and even uniform, actually yield similar estimation errors. See, e.g., \citet[p.~43]{Silverman86} and \citet[p.~12]{Racine07}. In our case, we choose the Gaussian kernel, i.e., the density of the standard $D$-variate normal distribution.

\noindent\textbf{Choice of Bandwidth.}

\noindent Unlike choice of kernel, the choice of bandwidth is of key importance in kernel regression (see generally \citealp{HardleEtAl88}). Initial kernel regression models treated the bandwidth parameter as scalar and constant over all observations in the dataset \citep{Parzen62,PriestleyChao72} and this still the dominant approach \citep[p.~15]{Racine07}. However, it leads to a significant bias if the density of any feature is highly nonuniform, and for multidimensional feature spaces it requires prior standardization of the feature scales.

Two most popular alternative approaches are the \emph{generalized nearest-neighbor bandwidth} \citep{LoftsgaardenQuesenberry65} and the \emph{adaptive nearest-neighbor bandwidth} \citep{BreimanEtAl77,Abramson82,Silverman86,Schucany95}:

\begin{definition}[Generalized Nearest-Neighbor Bandwidth]
    For $\mathbf{x}_{0}, \mathbf{x}_j \in \mathfrak{F}$, the \emph{generalized nearest-neighbor bandwidth} is given by:
    \begin{equation}
        \mathbf{h}_{\mathbf{x}_{0}, \mathbf{x}_j} = h_0 d(\mathbf{x}_{0}, \mathbf{x}_{N_k(\mathbf{x}_{0})}),
    \end{equation}
    where $d: \mathfrak{F} \times \mathfrak{F} \rightarrow \mathbb{R}_{\ge 0}$ is a metric, $N_k(\mathbf{x})$ is the index of the $k$-th nearest neighbor of $\mathbf{x}$ under $d$, and $h_0 \in \mathbb{R}_{+}$ is a scaling constant.
\end{definition}

\begin{definition}[Adaptive Nearest-Neighbor Bandwidth]
    For $\mathbf{x}_{0}, \mathbf{x}_j \in \mathfrak{F}$, the \emph{generalized nearest-neighbor bandwidth} is given by:
    \begin{equation}
        \mathbf{h}_{\mathbf{x}_{0}, \mathbf{x}_j} = h_0 d(\mathbf{x}_{0}, \mathbf{x}_{N_k(\mathbf{x}_{j})}),
    \end{equation}
    where $d: \mathfrak{F} \times \mathfrak{F} \rightarrow \mathbb{R}_{\ge 0}$ is a metric, $N_k(\mathbf{x})$ is the index of the $k$-th nearest neighbor of $\mathbf{x}$ under $d$, and $h_0 \in \mathbb{R}_{+}$ is a scaling constant.
\end{definition}

\begin{definition}[Adaptive Nearest-Neighbor Bandwidth]
    For $\mathbf{x}_{0}, \mathbf{x}_j \in \mathfrak{F}$, the \emph{adaptive nearest-neighbor bandwidth} is given by:
    \begin{equation}
        \mathbf{h}_{\mathbf{x}_{0}, \mathbf{x}_j} = h_0 d(\mathbf{x}_{0}, \mathbf{x}_{N_k(\mathbf{x}_{j})}),
    \end{equation}
    where $d: \mathfrak{F} \times \mathfrak{F} \rightarrow \mathbb{R}_{\ge 0}$ is a metric, $N_k(\mathbf{x})$ is the index of the $k$-th nearest neighbor of $\mathbf{x}$ under $d$, and $h_0 \in \mathbb{R}_{+}$ is a scaling constant.
\end{definition}

\noindent Generalized NN is more computationally efficient, since the nearest-neighbor search need only to be performed once per the point of estimation, while in adaptive NN it has to be performed for every realization of the predictor variables. On the other hand, adaptive NN yields smoother estimators -- generalized NN can result in non-differentiable peaks of the regression function. Motivated by the latter consideration, we use adaptive NN bandwidth, albeit in a modified multivariate version which enables us to have different bandwidths for different dimensions of the feature space:

\begin{definition}[Multivariate Adaptive Nearest-Neighbor Bandwidth]
    For $\mathbf{x}_{0}, \mathbf{x}_j \in \mathfrak{F}$, the $i$-th coordinate of the \emph{multivariate adaptive nearest-neighbor bandwidth}, $i = 1, \dots, D$, is given by:
    \begin{equation}
        (h_{\mathbf{x}_{0}, \mathbf{x}_j})_{i} = h_{0,i} |x_{0,i} - x_{N_k^i(\mathbf{x}_{j}),i}|,
    \end{equation}
    where $N_k^i(\mathbf{x})$ is the index of the $k$-th nearest neighbor of $\mathbf{x}$ along the $i$-th dimension of the feature space under the absolute difference metric, and $\mathbf{h}_0 \in \mathbb{R}_{+}^{D}$ is a scaling vector.
\end{definition}

The choice of a nearest-neighbor bandwidth requires us to choose additional hyperparameters of the model: the scaling vector $\mathbf{h}_0$ and the nearest-neighbor parameter $k$. This is usually done by leave-one-out cross-validation \citep{LiRacine04,HardleEtAl88} with the objective function defined either as an $L_1$ or $L_2$ distance between the predicted and actual value vectors \citep{CravenWahba78}, or as the Kullback-Leibler \citeyear{KullbackLeibler51} divergence between the former and the latter. We use the latter variant together with an optimization algorithm by \citet{HurvichEtAl98} which penalizes high-variance bandwidths (with variance measured as the trace of the parameter matrix) in a manner similar to the well-known Akaike information criterion \citep{Akaike74}.

\section{Measuring Deviation from the Seats-Votes Function}

By this point, we have estimated a party's expected seat share given its aggregate vote share and the competition patterns in the districts it contests. But what we actually need is a measure of how much the actual seat share deviates from that expectation. A natural choice would be the difference of the two. It is, however, inappropriate for two reasons:
\begin{description}
\item[First] Seat shares only assume values within a bounded interval $[0,1]$. Thus, in particular, if the expected seat share is different from $1/2$, the maximum deviations upwards and downwards differ.
\item[Second] There is no reason to expect seat share distributions to be even approximately symmetric around the mean, so deviation of $\varepsilon$ upwards might be significantly more or less probably than identical deviation downwards.
\end{description}

We therefore use another measure of deviation: the probability that a seat share deviating from the median more than the empirical seat share could have occurred randomly. Note how this quantity is analogous in definition to the $p$-value used in statistical hypothesis testing:

\begin{definition}{Electoral Bias $p$-Value}
    Let $s_i$ be an empirical seat share and let $\mu$ be the conditional distribution of the aggregate seat share given the empirical aggregate vote share and the empirical mean effective seat threshold, i.e., the value of the seats-votes function. Then the \emph{electoral bias $p$-value} is given by:
    \begin{equation}
        \pi_i = \min \{\mu((0, s_i)), \mu((s_i, 1))\} = \min \{F(s_i), 1 - F(s_i)\},
    \end{equation}
    where $F$ is the cumulative distribution function of $\mu$.
\end{definition}

We thus need not a regression estimator, but a conditional cumulative distribution function estimator. One approach would be to estimate the conditional density of $S_i$ \citep{Rosenblatt56,Parzen62} and integrate it numerically. This method, however, is prone to potential numerical errors. We therefore use another approach, relying on the fact that a conditional cumulative distribution function is defined in terms of the conditional expectation, and therefore the problem of estimating it can be treated as a special case of the kernel regression problem.


There remains one final problem: when comparing parties contesting different number of districts, we need an adjustment for the fact that the probability of getting an extreme value depends on that number (decreasing exponentially as the number of contested districts increases). In particular, except for very rare electoral ties, single-district parties always obtain extreme results. Thus, if for a party contesting $k$ districts we include parties contesting fewer districts in the training set, we overestimate the probability of obtaining an extreme seat share. To avoid that problem, the kernel model for parties with exactly $k$ districts, $k \in \mathbb{N}$, is trained only on parties with as many or fewer contested districts. If the distribution of the number of contested districts has a tail, it is optimal to adopt a cutoff point $k_0$ such that for the set $P_{c \ge k_0}$ of parties contesting $k_0$ or more districts each party is compared with a model trained on all parties in $P_{c \ge k_0}$.

\section{Aggregation}

The final step is the aggregation of party-level indices into a single election-level index of electoral bias. We would like our aggregation function to: (1) assign greater weight to major parties than to minor parties; (2) be sensitive to very low $p$-values and less sensitive to even substantial differences in large $p$-values; and (3) be comparable among elections, i.e., independent of the number of parties and districts. An easy example of such a function is the \emph{weighted geometric mean} given by:
\begin{equation}
    \pi := \exp \left(\sum_{i=1}^{n} w_i \log \pi_i \right),
\end{equation}
where $w_i$ is the number of votes cast for the $i$-th party divided by the number of all valid votes cast in the election (it differs from the aggregate vote share in that the denominator includes votes cast in districts not contested by the $i$-th party).

\section{Experimental Test}

Before applying our proposed method to empirical data, we wanted to sure that it really works -- both in terms of high precision (low number of false positives) and of high recall (low number of false negatives). But one fundamental problem in testing any method for the detection of gerrymandering, especially outside the familiar two-party pattern, lies in the fact that we have very few certain instances thereof. Therefore we first tested our method on a set of artificial (i.e., simulated) elections, consisting both of `fair' districting plans, drawn at random with a distribution intended to approximate the uniform distribution on the set of all admissible plans, and of `unfair' plans generated algorithmically. In order to ensure that voting patterns matched real-life elections, we used actual precinct-level data from the 2014 municipal election in our home city of Kraków. It was a multi-party election, but with two leading parties that were nearly tied in terms of votes. That allowed us to generate `gerrymandered' plans for both of them, improving the test quality.

\subsection{Experimental Setup}

Out baseline dataset consisted of a neighborhood graph of $452$ electoral precincts, each of which was assigned three parameters: precinct population, $w_k \in \mathbb{N}$, varying between 398 and 2926 (but with 90\% of the population taking values between 780 and 2420); party $p$'s vote share $p_k$, varying between $9.1\%$ and $66.7\%$ (but with 90\% of the population taking values between $20.8\%$ and $48.5\%$); and party $q$'s vote share $q_k$, varying between $11.5\%$ and $56.7\%$ (but with 90\% of the population taking values between $21.0\%$ and $44.1\%$). On the aggregate, party $p$ won the election with $33.15\%$ of the vote, but party $q$ was a close runner-up with $32.64\%$ of the vote. There have been seven third parties, but none of them had any chance of winning any seats (in particular, none has come first in any precinct). In drawing up plans, we fixed the number of districts at $43$ (the real-life number of seats in the municipal council) and the permissible population deviation at $25\%$.

As our training set, we used dataset $\mathcal{D}_{14}$, described in the following section.

\subsection{Algorithm for Generating Fair Plans}

Our sample of fair districting plans consisted of 128 partitions of the precinct graph generated using the Markov Chain Monte Carlo algorithm proposed by \citet{FifieldEtAl15}. It used the Swendsen-Wang algorithm \citep{SwendsenWang87}, as modified by \citet{BarbuZhu05}, to randomly walk the graph of solutions. In each iteration, we randomly `disable' some of the edges within each district of the starting districting plan (independently and with a fixed probability); identify connected components adjoining district boundaries; randomly choose $R$ such components (where $R$ is chosen from some fixed discrete distribution on $\mathbb{R}_{+}$) in such manner that they do not adjoin one another; identify admissible exchanges; and randomly accept or reject each such exchange using the Metropolis-Hastings criterion. \citet{BarbuZhu05} have shown that if $\Pr(R = 0) > 0$, the algorithm is ergodic, and \citet{FifieldEtAl15,FifieldEtAl20} -- that in such a case its stationary distribution is the uniform distribution on the set of admissible districtings. In practice this algorithm has a better rate of convergence than classical Metropolis-Hastings, but obtaining satisfactory performance still required additional heuristic optimizations like simulated annealing \citep{MarinariParisi92,GeyerThompson95}.

\subsection{Algorithm for Generating Unfair Plans}

To generate unfair districting plans we used an algorithm by Szufa et al. \citep[Ch.~3.7]{FlisEtAl23} based on integer linear programming. The essential idea is to consider all feasible districts (connected components of the precinct graph with aggregate population within the admissible district population range), $K_1, \dots, K_d$, and to solve the following optimization problem for party $x = p, q$:
\begin{problem}
    \textbf{For}
        $$\boldsymbol{\xi} \in \mathbb{B}^{d}$$
    \textbf{maximize}
    \begin{equation}
        \sum_{j = 1}^d \xi_j \sgn\left(\sum_{k \in K_j} x_k w_k - \sum_{k \in K_j} y_k w_k\right)
    \end{equation}
    \textbf{subject to}
    \begin{gather}
        \sum_{j=1}^d \xi_j = s, \\
        \sum_{j=1}^d \mathbf{1}_{K_j}(k) = 1 \textrm{ for every } k = 1, \dots, c,
    \end{gather}
    where $y = q$ if $x = p$ and $y = p$ if $x = q$.
\end{problem}

In other words, we choose such subset of feasible districts that maximizes the seat share of party $x$ subject to constraints that the number of chosen districts equals the number of seats $s$ and every precinct is assigned to exactly one district. Since this is a classical ILP problem, it can solved using a standard branch and bound algorithm.

In practice, it is infeasible to enumerate all possible districts with hundreds of precincts. We therefore first artificially combine leaf nodes, small precincts, and similar precincts until the number of precincts is reduced below 200. Only then we run the ILP algorithm and recover the full solution by replacing combined precincts with their original elements. Since the combining process can lead to suboptimality, we then run a local neighborhood search algorithm to find a local maximum.

The process as described finds an optimal ex post gerrymandering - in our case, we get a 36-seat districting plan for party $p$ and a 37-seat plan for party $q$. However, we can obtain less extreme instances of gerrymandering by including an additional constraint in our algorithm on the required margin of victory.

\subsection{Results}

Our sample of fair districting plans yielded a distribution of electoral results varying from $26$ to $17$ seats for party $p$, with the median at $21$. Those results corresponded to aggregate $p$-values between $.18$ and $.37$. Accordingly, none of the fair plans was classified as gerrymandered at the significance level $.05$, giving us perfect precision $1$.

Gerrymandered plans varied from a 37-seat to a 29-seat plan for party $q$, corresponding to aggregate $p$-values from $.006$ to $.064$, and from a 36-seat to a 29-seat plan for party $p$, corresponding to aggregate $p$-values from $.016$ to $.096$. In total, 24 out of 28 gerrymandered plans were classified as such at the significance level $.05$, yielding recall $.857$. Note, however, that all plans that we failed to recognize as gerrymandered were highly inefficient ones in terms of the seat benefit to the party doing gerrymandering.

\section{Empirical Test}

We have tested our method on data from four training sets of elections:
\begin{enumerate}
    \item $\mathcal{D}_{14}$, 2014 Polish municipal elections (this has been the case that originally motivated us to develop the method described in this paper) (2412 elections, 15,848 parties, 37,842 districts, 131,799 candidates),
    \item $\mathcal{D}_{18}$, 2018 Polish municipal elections (2145 elections, 10,302 parties, 32,173 districts, 86,479 candidates),
    \item $\mathcal{D}_{U}$, U.S. House of Representatives elections from the 1900-2016 period, where the election within each state is treated as a single election (2848 elections, 13,188 parties\footnotemark, 23,390 districts, 71,314 candidates),
    \item $\mathcal{D}_{N}$, national legislative elections from 15 countries (206 elections, 53,721 parties, 52,321 districts, 237,331 candidates).
\end{enumerate}
\footnotetext{We do not need to track party identity beyond any individual election, wherefore for instance the Republican party in, say, the 1994 House election in Pennsylvania and in the 1994 House election in New York (or the 1996 House election in Pennsylvania) is counted as two different parties. Hence the large number of parties in the U.S. election dataset.}

The following countries were included in the $\mathcal{D}_{N}$ dataset:
\begin{enumerate}
    \item United Kingdom -- all general elections from 1832 (47 cases); multi-member districts and Speakers running for reelection were dropped from the dataset;
    \item Canada -- all general elections from 1867 (42 cases);
    \item Denmark -- all general elections held under the FPTP system, i.e., those from 1849 to 1915 (32 cases);
    \item New Zealand -- all general elections from 1946 until introduction of MMP system in 1994 (17 cases);
    \item India -- all general elections from 1962 until 2014 (14 cases);
    \item Malaysia -- all general elections from 1959 until 2018 (12 cases);
    \item Philippines -- all general elections from 1987 until 1998, as well as single-member-district results from elections held under parallel voting from 1998 to 2013 (9 cases);
    \item Japan -- single-member-district results from elections held under parallel voting from 1996 to 2014 (7 cases);
    \item Ghana -- 2000-2016 elections (4 cases);
    \item South Africa -- 1984-1989 elections (4 cases);
    \item Poland -- upper house elections from 2011 (3 cases);
    \item Taiwan -- single-member-district results from elections held under parallel voting (3 cases);
    \item Nigeria -- 2003 and 2011 elections;
    \item Kenya -- 2002 and 2007 elections.
\end{enumerate}

Most of the national election data has been obtained from the Constituency-Level Election Archive \citep{dataCLEA}.

Those 15 countries were chosen as major countries using the FPTP system that have been categorized as at least partly free under the Freedom House Freedom in the World survey \citep{FreedomHouse22}. We have chosen elections that, even if not always democratic by modern standards, were at least minimally competitive (some opposition parties were able to field candidates and the results generally reliable). We note that the countries listed include cases with strong regional parties (UK, Canada) or very large number of small parties and independent candidates (India), as well as cases with party systems developing or otherwise fluid (19th century elections, developing country elections). They also include instances in which allegations of gerrymandering have already appeared in the literature \citep{HickmanKim92,Horiuchi04,McElwain08,Jou09,Brown05,Saravanamuttu18,Lau18,Ostwald19,IyerReddy13,Verma02,Christopher83}. For those reasons, we believe they constitute a good testing set.

Two Polish local election datasets were included as examples of extremely irregular competition patterns, especially the 2014 election, which has been the first one held under FPTP. By the 2018 election local party systems have somewhat settled, as evidenced by the smaller average number and effective number of candidates. The U.S. elections, on the other hand, were included to test whether the proposed method works well with regular two-party elections.

\begin{table}[htb]
\begin{tabularx}{\textwidth}{X|RRR|RR|RRR|RRR}
\toprule
\multicolumn{1}{c}{} & \multicolumn{3}{|c}{$\boldsymbol{c}$} & \multicolumn{2}{|c}{$\boldsymbol{n}$} & \multicolumn{3}{|c}{$\boldsymbol{\varphi}$} & \multicolumn{3}{|c}{$\boldsymbol{n_R}$} \\
\multicolumn{1}{c}{\multirow{-2}{*}{\textbf{dataset}}} & \multicolumn{1}{|c}{\textbf{med}} & \multicolumn{1}{c}{\textbf{min}} & \multicolumn{1}{c}{\textbf{max}} & \multicolumn{1}{|c}{\textbf{avg}} & \multicolumn{1}{c}{\textbf{max}} & \multicolumn{1}{|c}{\textbf{avg}} & \multicolumn{1}{c}{\textbf{min}} & \multicolumn{1}{c}{\textbf{max}} & \multicolumn{1}{|c}{\textbf{avg}} & \multicolumn{1}{c}{\textbf{min}} & \multicolumn{1}{c}{\textbf{max}} \\
\midrule
PL2014 & 15 & 1 & 23 & 3.48 & 10.22 & 2.78 & 1.55 & 6.79 & 3.29 & 2.00 & 7.20 \\
PL2018 & 15 & 1 & 15 & 2.81 & 6.60 & 2.37 & 1.62 & 4.72 & 2.74 & 1.91 & 5.60 \\
Intl & 197 & 8 & 659 & 3.78 & 21.21 & 2.24 & 1.65 & 3.62 & 2.59 & 1.91 & 4.32 \\
US & 6 & 1 & 53 & 2.96 & 9.00 & 1.87 & 1.00 & 3.82 & 2.06 & 1.00 & 4.33 \\
\bottomrule
\end{tabularx}
\label{tbl:datasets}
\caption{Basic characteristics of electoral datasets: the number of districts $c$, the number of candidates $n$ averaged over districts, the effective number of candidates $\varphi$ averaged over districts, and the number of candidates with at least $5\%$ vote share averaged over districts. We omit the $\min$ column for $n$, as it had value $2.00$ for all datasets. Unopposed districts have been dropped before any computations.}
\end{table}

\subsection{Results by Training Set}

In this subsection, we report the raw results for our training sets and test whether the method agrees with classical measures of gerrymandering for two-party U.S. elections.

\begin{figure}[htb]
    \centering
    \includegraphics[width=0.49\textwidth]{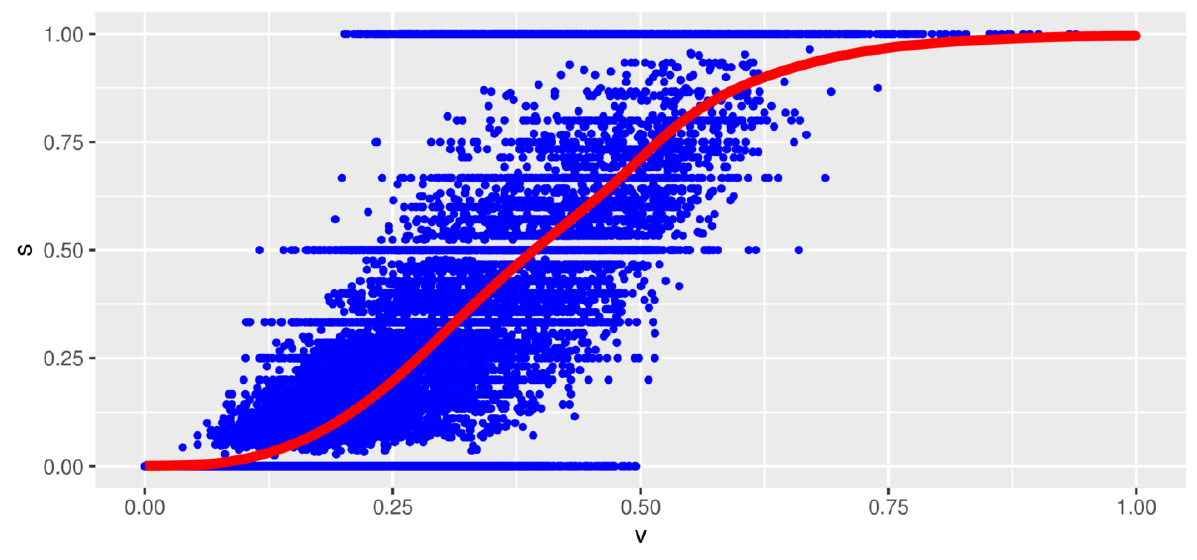}
    \includegraphics[width=0.49\textwidth]{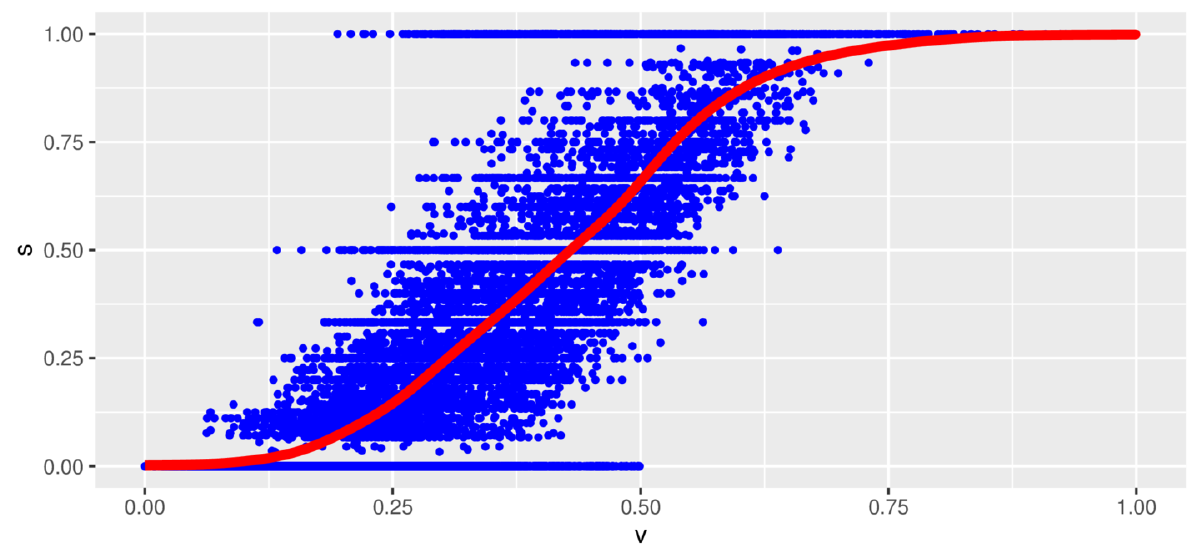}
    \includegraphics[width=0.49\textwidth]{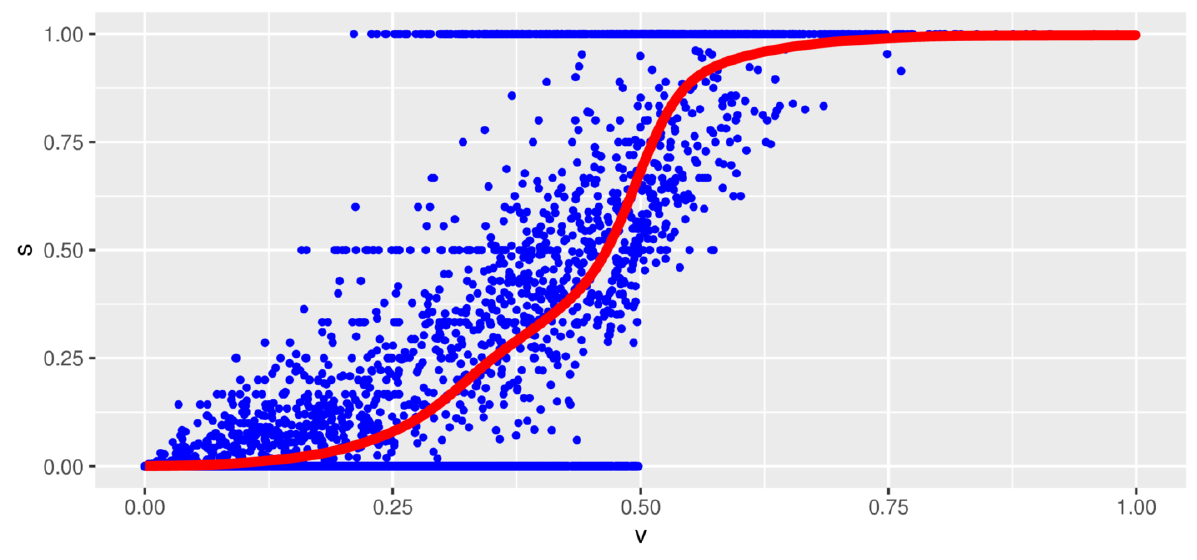}
    \includegraphics[width=0.49\textwidth]{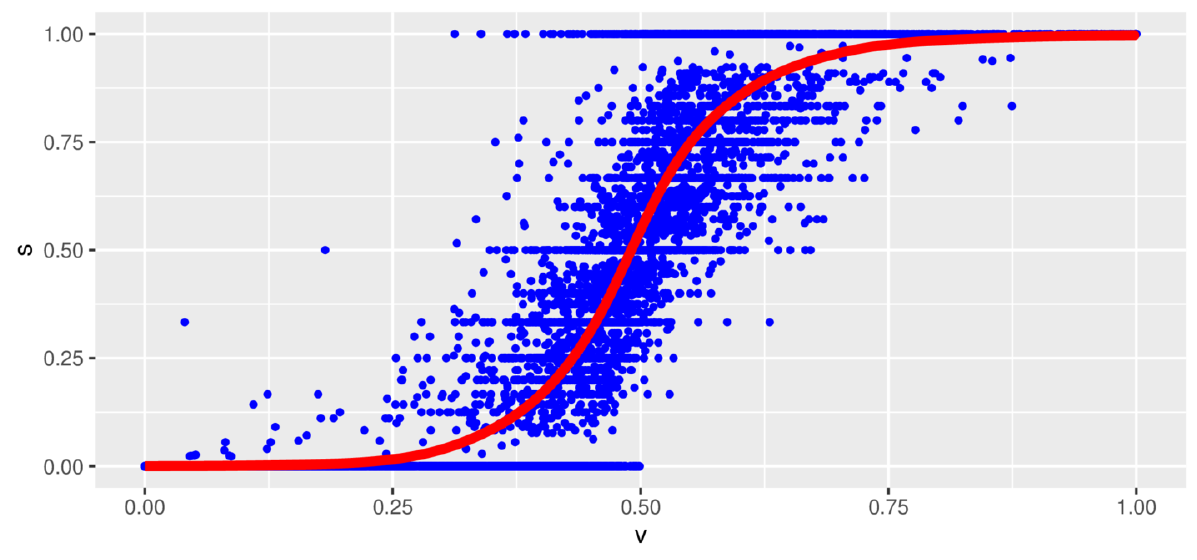}
    \caption{Nonparametric seats-votes curves for the four training sets (from upper left: $D_{14}$, $D_{18}$, $D_N$, $D_U$).}
    \label{fig:svPlot}
\end{figure}

\begin{figure}[htb]
    \centering
    \includegraphics[width=\textwidth]{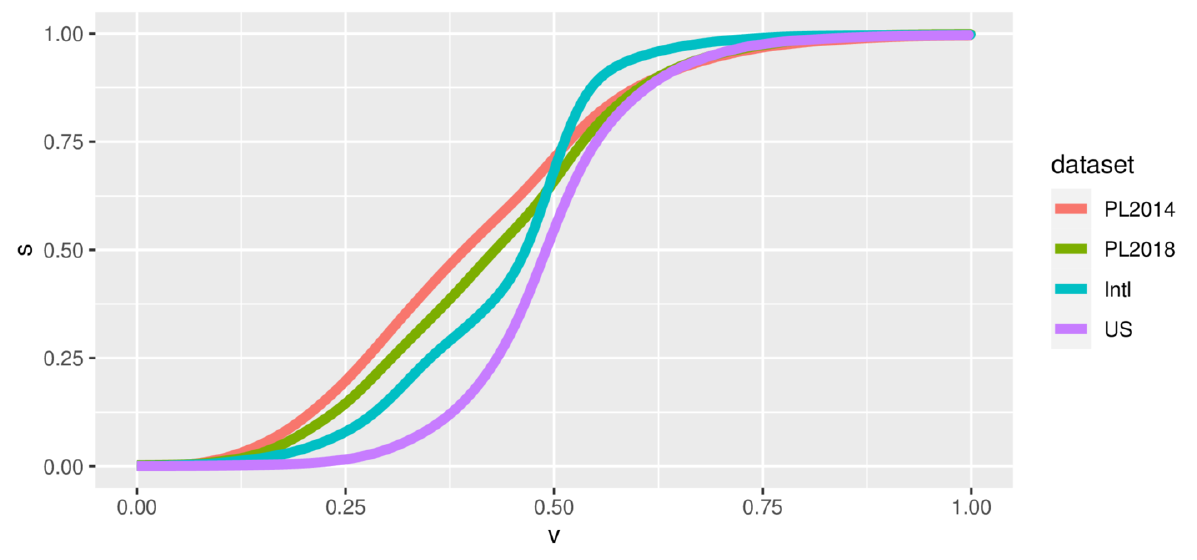}
    \caption{Nonparametric seats-votes curves for the four training sets -- comparison.}
    \label{fig:svCurves}
\end{figure}

\begin{figure}[htb]
    \centering
    \includegraphics[width=\textwidth]{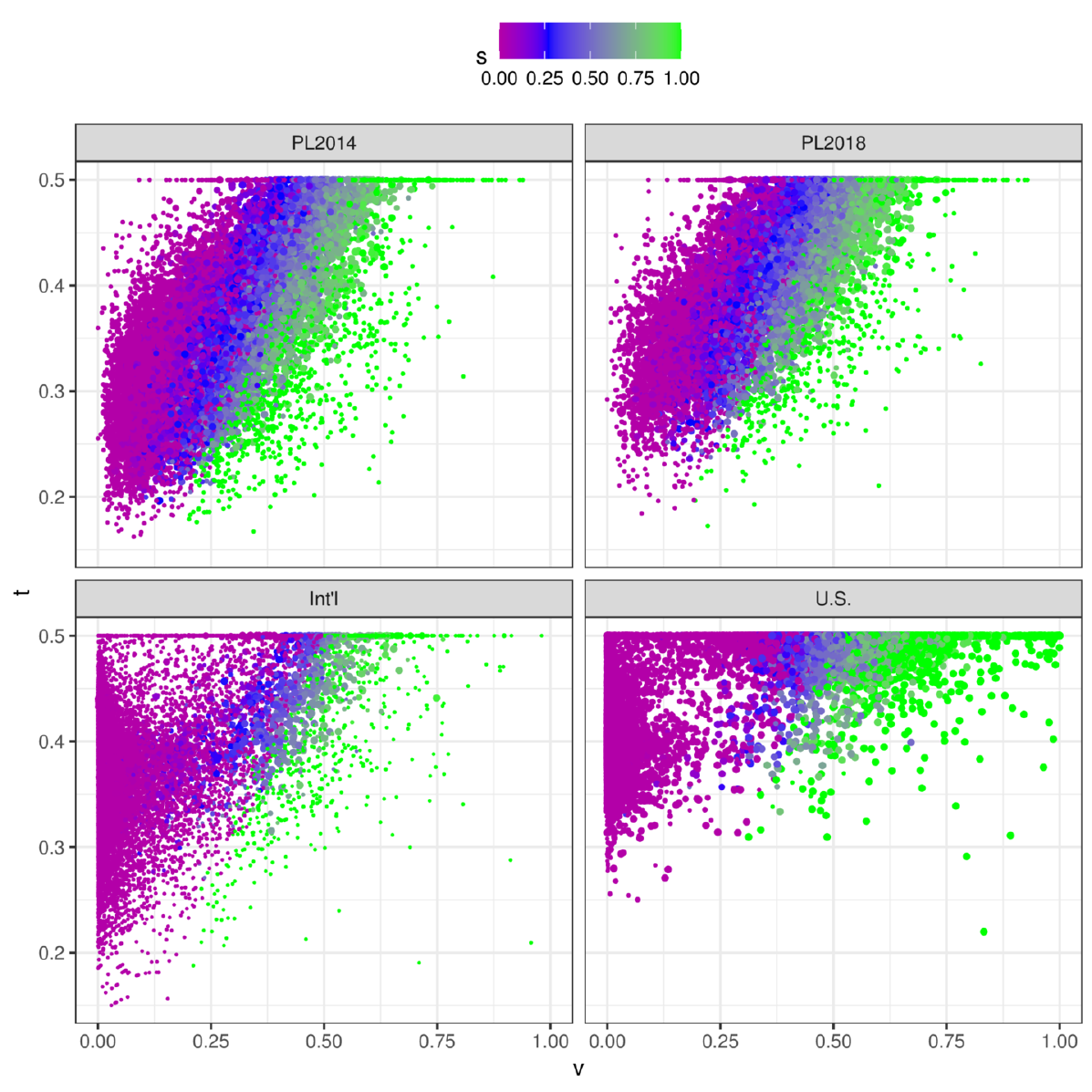}
    \caption{The empirical seat share as a function of the vote share and the effective seat threshold.}
    \label{fig:svtPlotEmpirical}
\end{figure}

\begin{figure}[htb]
    \centering
    \includegraphics[width=\textwidth]{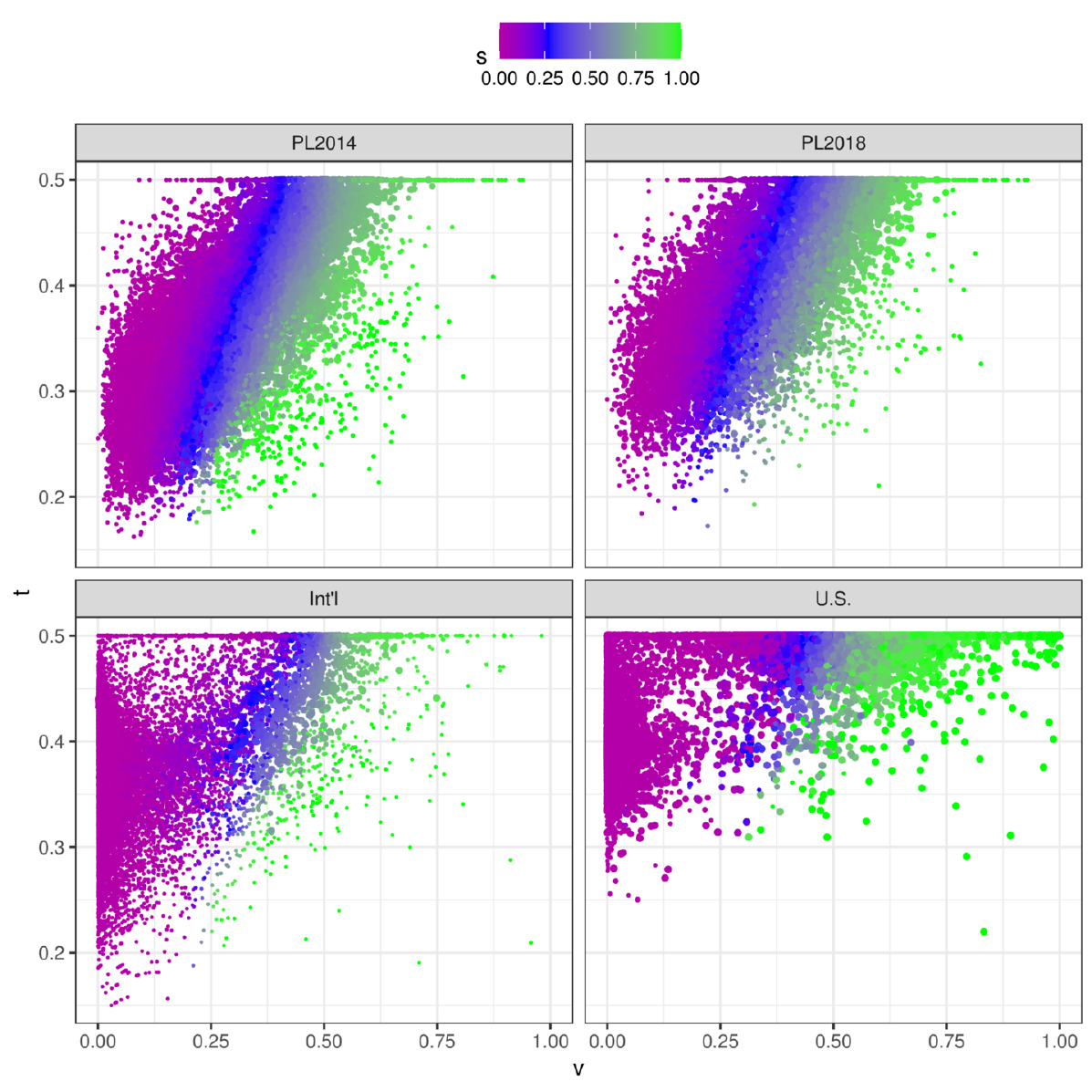}
    \caption{The expected seat share as a function of the vote share and the effective seat threshold.}
    \label{fig:svtPlotExpected}
\end{figure}

\begin{figure}[htb]
    \centering
    \includegraphics[width=\textwidth]{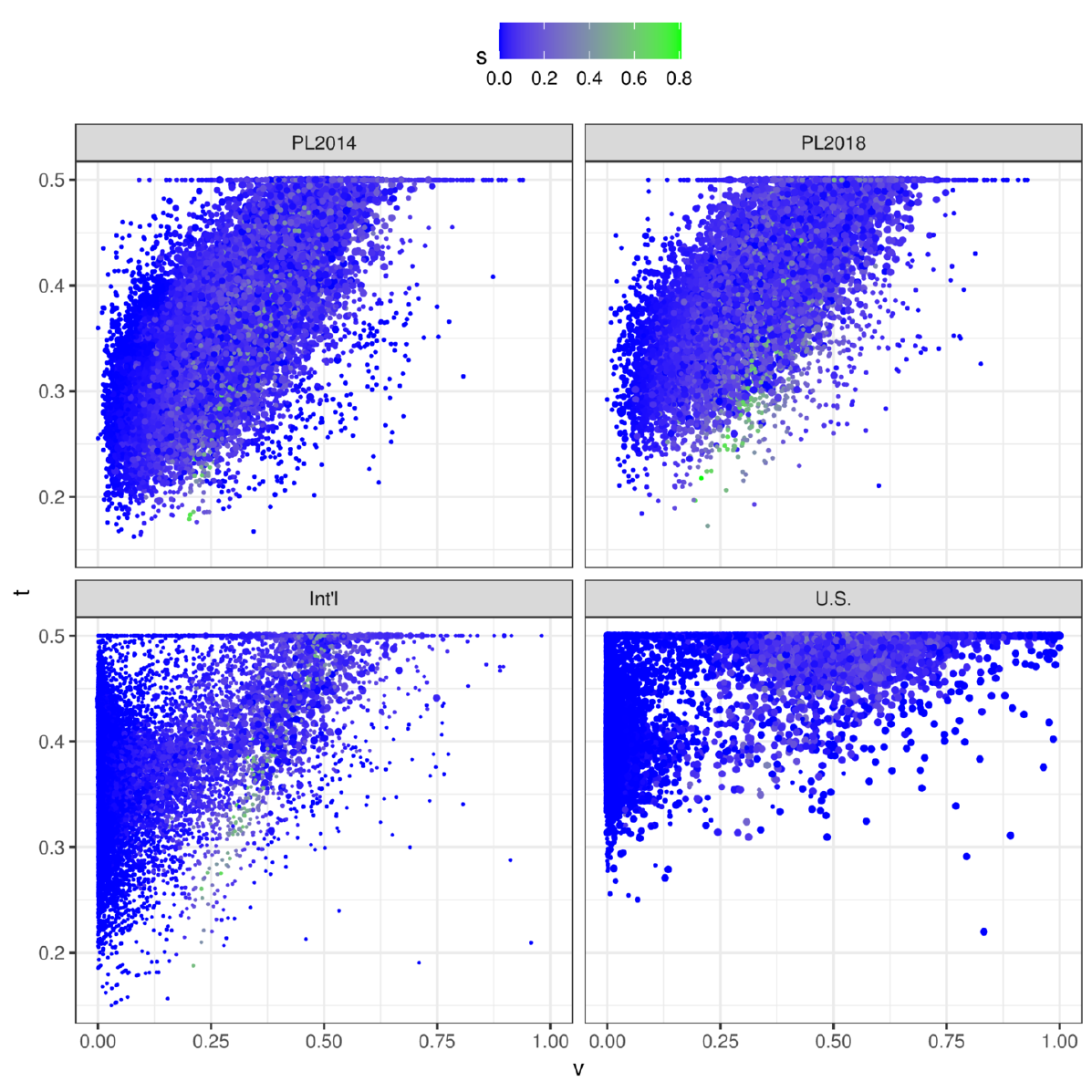}
    \caption{Difference between the empirical and expected seat share as a function of the vote share and the effective seat threshold.}
    \label{fig:svtPlotError}
\end{figure}

The incidence of electoral bias in the four datasets under consideration was as follows:

\begin{table}[H]
    \centering
    \begin{tabularx}{0.8\textwidth}{lRR}
    \toprule
        dataset & average $p$-value & $\Pr(\pi < .05)$ \\
    \midrule
        $\mathcal{D}_{14}$ (Poland 2014) & $.240$ & $.0079$ \\
        $\mathcal{D}_{18}$ (Poland 2018) & $.241$ & $.0141$ \\
        $\mathcal{D}_N$ (non-U.S. national elections) & $.311$ & $.0049$ \\
        $\mathcal{D}_U$ (U.S. elections) & $.307$ & $.0021$ \\
        $\mathcal{D}_U$ restricted to post-1970 elections & $.268$ & $.0122$ \\
    \bottomrule
    \end{tabularx}
    \caption{Incidence of Electoral Bias in Four Election Datasets.}
    \label{tab:incidBias}
\end{table}

Finding an appropriate baseline for comparison, however, is quite difficult. If we were to assume that that party $p$-values behave like statistical $p$-values, i.e., follow the uniform distribution on $(0,1)$, and that they are independent, we would expect the election $p$-values to follow an absolutely continuous distribution with density given by:
\begin{equation}
    f(x) = \frac{(-2\log x)^{n-1}}{2x(n-1)!}
\end{equation}
for $x \in (0, 1)$. However, while the first of those assumptions is realistic, the second is decidedly not: electoral bias is a zero-sum phenomenon and if an election is biased in favor of some set of parties, it must necessarily be biased against another set.

But even if we were able to easily model the expected scale of random electoral bias, it would still be impossible to determine whether deviation from it is caused by the deficiencies of our method or by actual instances of gerrymandering. Hence, a more appropriate test would be to analyze whether our measure agrees with other methods for detecting gerrymandering used in the literature. Of course, we can do this test only for two-party elections, such as those from the U.S. elections dataset. As many modern methods tend to assume the absence of large-scale malapportionment, we have dropped the pre-1970 observations. We have compared our coefficient with absolute values of four classical indices: Gelman-King partisan bias, efficiency gap, mean-median difference, and declination coefficient. Scores for those methods were obtained from PlanScore \citep{GreenwoodEtAl23}. Expected association is negative, since for all classical methods high values of the index are indicative of gerrymandering, while for our method values close to $0$ indicate electoral bias.

\begin{table}[H]
    \centering
    \begin{tabularx}{0.8\textwidth}{lRRRRl}
    \toprule
        Coefficient & Estimate & Std. Error & t Value & $p$-value & \\
    \midrule
        (Intercept) & .340 & .011 & 29.785 & < 2e-16 & *** \\
        abs(bias) & .278 & .186 & 1.492 & 0.13627 & \\
        abs(effGap) & -.895 & .189 & -4.725 & 2.99e-06 & *** \\
        abs(meanMed) & -.781 & .278 & -2.807 & 0.00519 & ** \\
        abs(declin) & .116 & .060 & 1.923 & 0.05500 & . \\
    \bottomrule
    \end{tabularx}
    \caption{Correlation with Classical Measures of Gerrymandering.}
    \label{tab:corelations}
\end{table}

For other datasets, we are at present left with qualitative analysis of the most biased cases. For the full U.S. dataset $D_U$ these were: several Missouri elections from the 1900s to 1920s (especially the 1926, 1916, 1906, and 1902), the 1934 Indiana and New Jersey elections, and the 1934 New Jersey elections. Missouri and Indiana were at the time highly malapportioned \citep{DavidEisenberg61}, but the case of New Jersey requires further study. Nevertheless, an inspection of the results suggests that something was definitely amiss in all of those elections: while two-party vote shares were very close to $1/2$, the discrepancy of seat-shares (i.e., the partisan bias) was quite high, e.g., $.75$ to $.25$ in Missouri in 1926

For the non-U.S. national elections dataset the most biased instances include the 1874 U.K. general election (a famous electoral inversion where the Liberal Party decisively lost in terms of seats -- 242 to 350 -- despite winning a plurality of the popular vote), the 1873 Danish Folketing election (held in highly malapportioned districts), the 1882 Canadian federal election, and the 1841 U.K. general election, while among modern elections -- the 2013 Malaysian election, the 2014 Indian general election, the 2005 and 2009 Japanese elections, and the 1983 U.K. general elections (with the non-intentional pro-Labour and anti-SDP-Liberal Alliance bias resulting from geographical patterns documented by \citep{RossiterEtAl97,RossiterEtAl99,JohnstonEtAl98,Blau02}).

\singlespacing
\bibliographystyle{plainnat}
\bibliography{Gerrymandering}

\end{document}